# Increasing Papers' Discoverability with Precise Semantic Labeling: the sci.AI Platform.

Roman GURINOVICH [a,1], Alexander PASHUK [a], Yuriy PETROVSKIY [a,b], Alex DMITRIEVSKIJ [a], Oleg KURYAN [a], Alexei SCERBACOV [a], Antonia TIGGRE [a], Elena MOROZ [a,c], Yuri NIKOLSKY [a]

[a] *sci.AI, Xpansa, Mustamae Tee 5, 10616, Tallinn, Estonia*
[b] *Odessa National Medical University, Valikhovs'kyi Ln, 2, Odesa, Ukraine, 65000*
[c] *IITP RAS, Bolshoy Karetny per. 19, build.1, Moscow 127051 Russia*

**Abstract.** The number of published findings in biomedicine increases continually. At the same time, specifics of the domain's terminology complicates the task of relevant publications retrieval. In the current research, we investigate influence of terms' variability and ambiguity on a paper's likelihood of being retrieved. We obtained statistics that demonstrate significance of the issue and its challenges, followed by presenting the sci.AI platform, which allows precise terms labeling as a resolution.

1. **Key objectives of the study and its significance**

Over the last two decades, life sciences articles have become substantially more complex, reflecting technological evolution, particularly OMICs experimentation, increasing cooperation between multiple institutions, and involving more advanced math and statistics applied to the data. In many publications, plain unstructured text is supported by algorithms, code, and multiple files of processed and raw datasets with annotated metadata and graphs. With such enhancements in place, experimental articles, per se, might become a driving force of the Literature Based Discoveries (LBD) [1]. Recently, the whole field of "meta-analysis" has arose to describe "dry lab" studies on normalization, unification, and analysis of many similar datasets derived from different labs and projects. However, a number of experimental papers are missed, because they cannot be retrieved from the body of literature by keywords search. Needless to say that scientists are keenly interested in higher discoverability of their published research and referencing to their findings, as citation index becomes an increasingly prevalent metrics in evaluation of their work. The issue can be addressed with proper semantic labeling of the texts as the very first step in global analysis of the research reports.

---

[1] Corresponding Author: roman.gurinovich@xpansa.com

The state-of-the-art section reflects the ways published papers are algorithmically processed in text mining applications.  Such computations rely on preexisting, statistically supported information, while text mining of the scientific literature targets novel findings. This leads to Information Retrieval (IR) and then Information Extraction (IE) underperformance when applied to scientific literature.

The objective of the paper is to consider just one issue of many in biomedical texts processing: false terms recognition caused by ambiguity of the concepts' names and multiple-terms spelling variants. This leads to at least two undesirable effects:

1. Lower recall rate when search engines and aggregators retrieve articles, so the target audience does not receive a full set of relevant papers.

2. Retrieving a paper that is irrelevant to the sought-for concept. For example, reader can query 'cat' with the 'cat' animal in mind but receive texts about the 'CAT' gene.

We address:

a. A global need of initial transformation of the plain text to a machine-readable format; and

b. Uncertainty issue mentioned above;

by releasing the sci.AI system. This system combines automatic metatagging and manual validation of the results by the author or reader and supports generating semantic structures during writing and editorial processing.  Human validation eliminates almost any possibility of term misinterpretation in the following IR and IE tasks, because authors can be expected to have a comprehensive understanding of the concepts being mentioned in their papers and can supervise the machine's results.

**2. State of the art in the field of biomedical texts semanticization**

Computational Linguistics is one of the most dynamic fields with innovations being released almost monthly. Unfortunately, there is no solid state-of-the-art solution for biomedical text labeling yet that unites all the latest advances in each subfield into a single package.

The first subfield is metatagging standards and paradigms. Semantic Web's objects, concepts, knowledge association, and data representation utilize schema.org vocabularies and W3C RDF/XML [2]. A current limitation is the lack of a similar single schema for the life sciences. Former related initiatives here are W3C Scholarly HTML [11] and JATS4R [12]. Still both schemas do not provide a standard namespace for biomedical concepts labeling.

The second subfield is terms labeling or Named Entities Recognition.

Just as in many other areas, deep learning and neural networks (NN) methods are increasingly popular for extracting information from professional texts [7, 8]. NN algorithms are rather generic and can be applied for the text analysis in unsupervised fashion (i.e., to a variety of texts without establishing prior rules generation). However, its precision and recall hardy depend on statistical data and cannot be considered as stable solution for concepts and challenges that have appeared recently. Still, NN demonstrates the highest recognition rates [16, 17] among automatic methods.

Then there are methods of increasing precision by reducing concepts ambiguity by connecting the same concepts in various ontologies.

UMLS (Unified Medical Language System), in combination with MetaMap, provide graph-like links between objects from various ontologies, as widely-accepted solution for the Word Sense Disambiguation (WSD) task in the biomedical domain. Essentially, UMLS represents a metaontology of biomedical terms and concepts. UMLS is extensive and well supported by NIH, and it is in constant development. Future considerations include possibly connecting this data to the sci.AI application. Currently, there are several limitations:

1. Lack of details for specialised ontologies, such as Uniprot and ChEBI.

2. Focus on the indexing task for the NCBI. This leads to the same dropdown in precision and recall of post-publication text processing.

3. It is not a simple plug-and-play solution for the publishing industry [6].

SciGraph by the Neo4j [13] framework allows objects to be interconnected and can be used as a technical basis for future metaontologies.

## 3. Design and Methodology

Resolving terms' ambiguity and variability represents a significant challenge in text processing. Here, we investigated how these factors affect the paper's influence. Such causality is assumed based on the logic that findings described in the paper can be reused and cited—only if the paper will be discovered by the readers first. To model paper's influence potential mathematically, we defined Paper's Influence as a function of a variable we called Discoverability.

$$Paper's\ Influence\ =\ f(Paper's\ Discoverability) \qquad (1)$$

The paper's potential for influence greatly depends on how accurately search engines and aggregators solve the IR task. For further explanation, we will continue with our query scenario from above. A reader is discovering the paper about animal 'cat' after querying string $Q_i$ 'genes of cat', including term $t_{ij}$ 'cat' while meaning biomedical concept $c_i$ 'Felis catus', corresponding to object '9685' in the ontology [14]. Concept $c_i$ then can be referenced with any term (spelling variant) $t_{ij}$ of the set $T_i$

$$T_i\ =\ \{t_{i1}, t_{i2}, \ldots, t_{iN}\} \in c_i \qquad (2)$$

If we assume that a reader will read the paper if the search engine returned it in response to the query $Q_i$, then "discoverability" is a synonym of "retrieval". We can then apply two major IR metrics, recall and precision:

$$recall\ =\ \frac{number\ of\ relevant\ papers\ retrieved}{number\ of\ all\ existing\ papers\ about\ the\ concept}\ = \qquad (3)$$
$$=\ P(retrieved\ papers\ \supseteq T_i\ |\ existing\ relevant\ papers\ \supseteq c_i)$$

$$precision\ =\ \frac{number\ of\ relevant\ papers\ retrieved}{number\ of\ retrieved\ papers}\ = \qquad (4)$$
$$=\ P(relevant\ papers\ among\ retrieved\ \supseteq c_i\ |\ retrieved\ papers\ \supseteq T_i)$$

As long as such cases could be found across biomedical terminology, when concept can have several synonyms (variable terms) or single term can refer to several concepts (ambiguous terms), probabilistic precision and recall can be calculated based on the numbers of possible outcomes when querying $Q_i \supseteq t_{ij}$. For example, texts "TNF alpha", "TNFa" and "TNF α" are variants of the object Uniprot [P01375]. This means that if search engine was queried with "TNF alpha", an ideal result would return all documents that contain all three variants. Still, due to existence of the several variants, the there is a probability ≥ 0 that some of them will not be considered.

We can estimate chances of such event using a basic definition of the probability as the ratio of the number of favorable outcomes to the total number of possible outcomes. Term's ambiguity and variability define those numbers of possible outcomes. Finally, when we know precision and recall of the paper's retrieving while searching for the concept $c_i$, we can answer specific questions about discoverability of the paper in some kind of progression order.

Question 1. How many papers out of all existing literature about concept $c_i$ can be retrieved, when there is set $T_i$, all terms of which refer to this concept $c_i$ ?

$$
\begin{aligned}
&recall(variability) \\
&= \frac{number\ of\ retrieved\ papers\ with\ t_{i1} + \ldots + number\ of\ retrieved\ papers\ with\ t_{iN}}{total\ number\ of\ relevant\ existing\ papers\ about\ c_i} = \\
&= \frac{\Sigma\ retrieved\ papers\ with\ T_i}{total\ \Sigma\ papers\ relevant\ c_i}
\end{aligned}
\qquad (5)
$$

Question 2. How many papers out of retrieved and containing terms from $T_i$ mention concept $c_i$ specifically? As long as only recorded synonyms are proved to exist, we can assume that all synonyms from the ontology and generated variants constitute a full dictionary of the concept, and

$$
\begin{aligned}
&precision(variability) \\
&= \frac{number\ of\ relevant\ papers\ about\ c_i}{number\ of\ retrieved\ papers\ with\ t_{i1} + \ldots + number\ retrieved\ papers\ with\ t_{iN}} = \\
&= \frac{\Sigma\ relevant\ out\ of\ retrieved\ papers\ about\ c_i}{\Sigma\ retrieved\ papers\ with\ T_i}, \\
&\forall t_{ij}, T_i \in c_i, \\
&\Rightarrow \Sigma\ relevant\ papers\ with\ c_i = \Sigma\ retrieved\ papers\ with\ T_i \Rightarrow \\
&\Rightarrow precision(variability) = 1
\end{aligned}
\qquad (6)
$$

This means that precision for the specific concept does not depend on the number of variants, as long as we assume that all variants are describing the same concept in the event.

Question 3. There is term $t_{ij}$ which refers to the concept $c_i$ or another concept $c_m$. How many papers out of retrieved and containing term $t_{ij}$, are talking about the concept $c_i$ exactly?

$$
\begin{aligned}
&precision(ambiguity) = \\
&= \frac{number\ of\ relevant\ papers\ about\ c_i}{number\ of\ retrieved\ papers\ about\ c_i + \ldots + number\ of existing\ papers\ about\ c_m} =
\end{aligned}
\qquad (7)
$$

$$= \frac{\Sigma \text{ relevant out of retrieved papers about } c_i}{\Sigma \text{ retrieved papers } c_i \cap \ldots \cap c_m = \{t_{ij}\}}$$

Question 4. Did we receive all papers containing term $t_{ij}$? (Answer: Yes, obviously. Continuing to ask this question is important for keeping track of the general recall and precision derivation)

$$recall(ambiguity) = \frac{\text{number of retrieved papers with } t_{ij}}{\text{total number of relevant existing papers with } t_{ij}} = 1 \qquad (8)$$

Question 5. What is the overall probability of retrieving a relevant paper for the concept $c_i$ that has many variants $\{t_{i1}, t_{i2}, \ldots, t_{iN}\}$ and some of them $c_i \cap \ldots \cap c_m = \{t_{ij}\}$ are also found in the other concepts?

This means a probability of two independent events: A = the concept has spelling variants, and B = those variants can be found in several concepts. Therefore P (A and B) will be multiples of the probabilities above:

$$recall(variability \text{ and } ambiguity) = recall(variability) * recall(ambiguity) =$$
$$= \frac{\Sigma \text{ retrieved papers with } T_i}{\text{total } \Sigma \text{ papers relevant } c_i} * 1 \qquad (9)$$

$$precision(variability \text{ and } ambiguity) =$$
$$= precision(variability) * precision(ambiguity) =$$
$$= \frac{\Sigma \text{ relevant out of retrieved papers about } c_i}{\Sigma \text{ retrieved papers } c_i \cap \ldots \cap c_m = \{t_{ij}\}} * 1 \qquad (10)$$

If operating only with the number of variants per concept, then prior probability of variants occurrence can be approximated as uniformly distributed, as long as actual frequency of terms occurrence in the papers will be retrieved in the next steps. This means that, in the first approximation, occurrence = {True, False} of the term can be sufficient variable to estimate the minimum expected probabilities:

$$prior\ recall(variability \text{ and } ambiguity) = \frac{1}{\text{Number of terms variants per concept}} \qquad (11)$$

$$prior\ precision(variability \text{ and } ambiguity) =$$
$$= \frac{1}{\text{Number of concepts per single term variant}} \qquad (12)$$

As long as the Number of the terms per concept $\geq 1$ (each object has at least a main name) and Number of concepts per single term variant $\geq 1$ (each term is related to at least one object) and they are in the denominator of the retrieving probabilities above—terms' variability and ambiguity will always reduce (at least, will not increase) recall and precision, respectively, when searching for the paper.

In order to estimate influence of the existing terms' uncertainty on the papers discoverability, we have searched for:

1. homographs across Uniprot, ICD-10, ChEBI, MeSH, Drugbank, and Gene Ontology databases;

2. possible spelling variants for the same objects;
3. actually used terms' variants in the 26782464 Pubmed, 26404 Bioline and 5426 eLife papers.

MeSH Categories G–Z were not analysed because they contain generic objects, such as countries' names, which are out of scope of sci.AI semanticization for now.

Our research is ongoing and the latest results can be found on the sci.AI webpage [20].

**Table 1.** Variability in the ontologies and influence on paper's recall

| Ontology name | Ontology size, n of IDs | Number of synonyms | Average number of synonyms | Number of synonyms that were found in the papers | Number of synonyms with variants | Average number of synonyms with variants | Number of variants that were found in the papers | Maximum number of found variants per concept | The smallest expected recall |
|---|---|---|---|---|---|---|---|---|---|
| Uniprot | 553667 | 1018837 | 1.8402 | 702942 | 1729359 | 3.1235 | 936673 | 84 | 0.0119 |
| ChEBI | 104854 | 201061 | 1.9175 | 175750 | 510817 | 4.8723 | 67159 | 81 | 0.0123 |
| Gene Ontology | 46517 | 173156 | 3.7224 | 47073 | 287903 | 6.1914 | 54711 | 94 | 0.0106 |
| Drugbank | 8221 | 28980 | 3.5251 | 15724 | 154704 | 18.8204 | 28445 | 243 | 0.0041 |
| ICD-10 | 11420 | 20728 | 1.8150 | 9680 | 30463 | 2.6675 | 14883 | 19 | 0.0526 |
| MeSH (A-F tree) | 23716 | 199486 | 8.4114 | 139078 | 309189 | 13.0459 | 170898 | 210 | 0.0048 |

We had not only considered synonyms that exist in the ontologies but also created a rules-based term variant generator (TVG) to cover a case when the same object, Uniprot [P01375], might be written as "TNF alpha", "TNFa", or "TNF α" in a paper. Next generating techniques groups were utilized:
- orthographic;
- abbreviations and acronyms;
- inflectional variations;
- morphological variations;
- structural recombinations [4, 5, 6].

Table 1 shows average number of original terms' synonyms and how much variants were generated. Then we've searched for them in the papers. There is increase of the concept detection of 2.03 - 3 times more when searching for all variants.

Table 2 shows how much objects has terms with identical spellings, i.e. ambiguous terms. Higher overlap within the same ontology than across other ontologies makes algorithmic recognition even more challenging tasks, because algorithms have to distinguish objects within the same class.

**Table 2.** Ambiguity in the ontologies and influence on paper's precision

| Ontology name | Ontology size, n of IDs | Number of the objects' names with the same spelling within ontology | Number of the objects with the same spelling for all variants *within ontology* | Number of the objects with the same spelling for all variants *across other ontologies* | Number of the objects with the same spelling for all variants across *the same and other ontologies* | Maximum number of objects with the same spelling | The smallest expected precision |
|---|---|---|---|---|---|---|---|
| Uniprot | 553667 | 0 | 75271 | 3426 | 78697 | 17 | 0.0588 |
| ChEBI | 104854 | 0 | 14921 | 597 | 15518 | 1780 | 0.0006 |
| Gene Ontology | 46517 | 0 | 2882 | 1177 | 4059 | 9 | 0.1111 |
| Drugbank | 8221 | 0 | 20771 | 478 | 21249 | 766 | 0.0013 |
| ICD-10 | 11420 | 0 | 1882 | 762 | 2644 | 6 | 0.1667 |
| MeSH (A-F tree) | 23716 | - | - | - | - | - | - |

Fig.1 shows overall influence of the variability and ambiguity of the terminology on paper's discoverability.

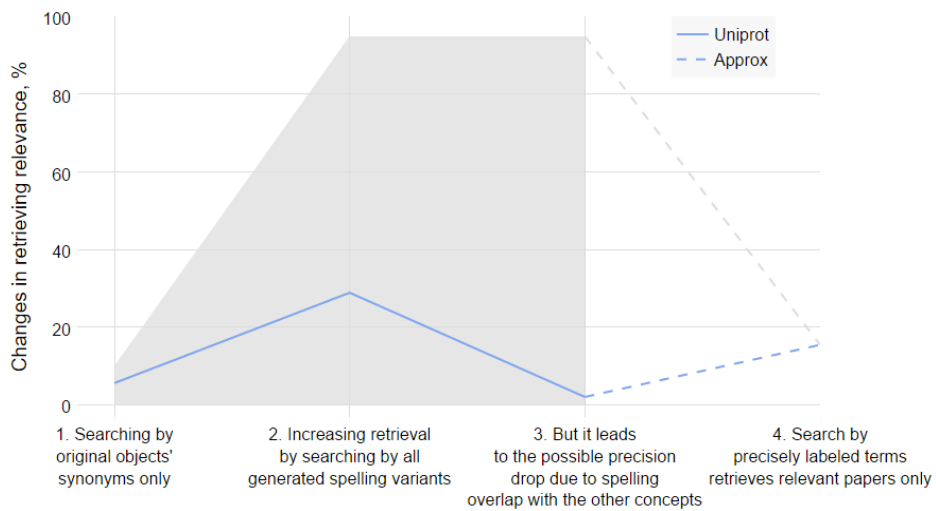

a) Uniprot

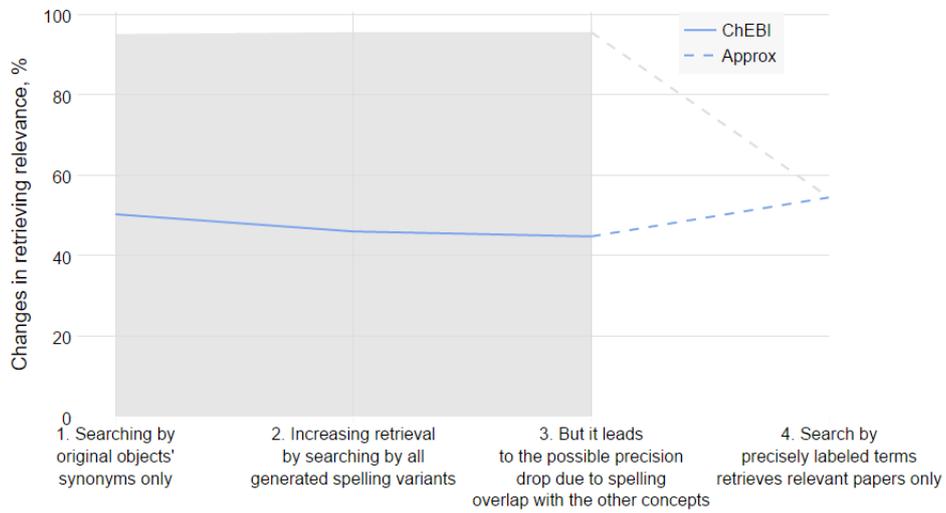

b) ChEBI

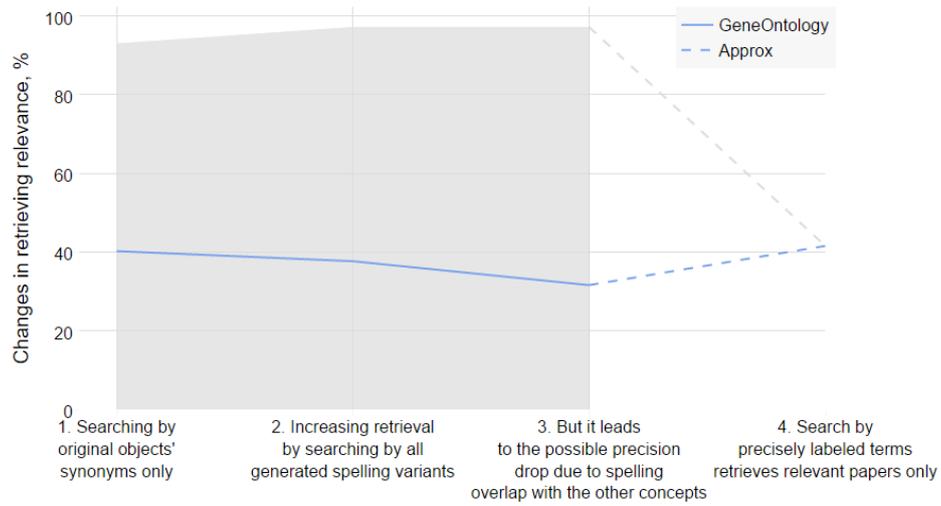

c) GeneOntology

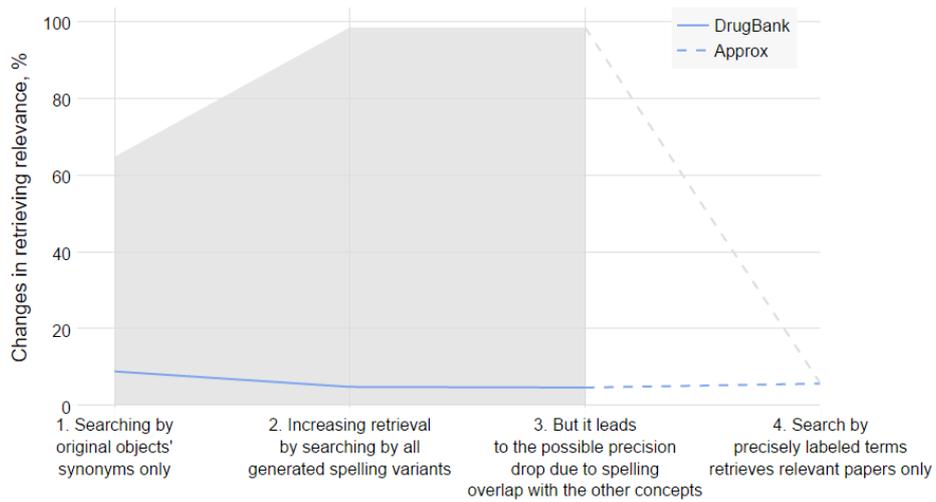

d) DrugBank

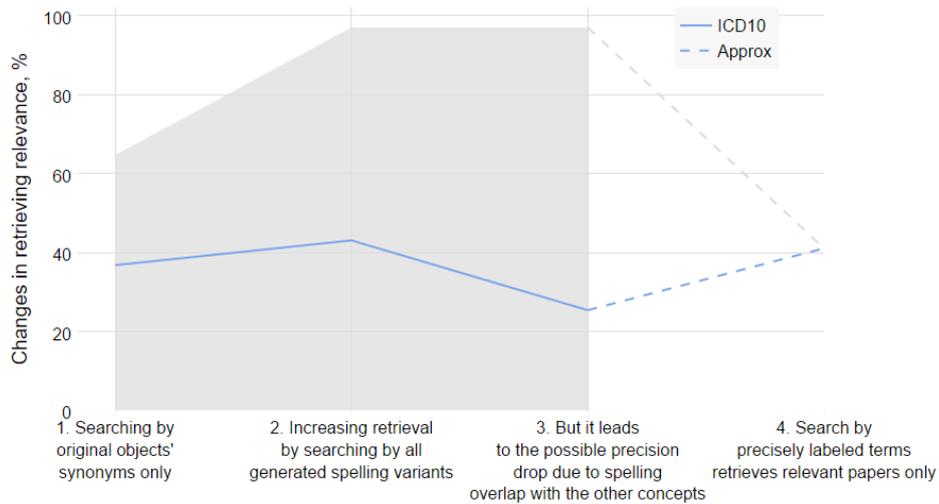

e) ICD10

**Figure 1.** Influence of the terminology variability and ambiguity on a paper retrieval when text contains proteins, chemical elements, genes, drugs, diseases.

When searching by original synonyms only, average likelihood of finding papers is lower than searching by all possible variants. Retrieving higher amount of the papers can be done at the cost of their relevance. Increasing amount of variants leads to the drop of the probabilistic precision. Relevance can be guaranteed only in case of labeling terms and searching by exact ID instead of a string. As long as current literature is not labeled, exact recall and precision can't be calculated for. We used relative changes instead, to visualize scale of the issue across most of the available literature.

Fig.2 shows the distribution of the variability across ontologies.

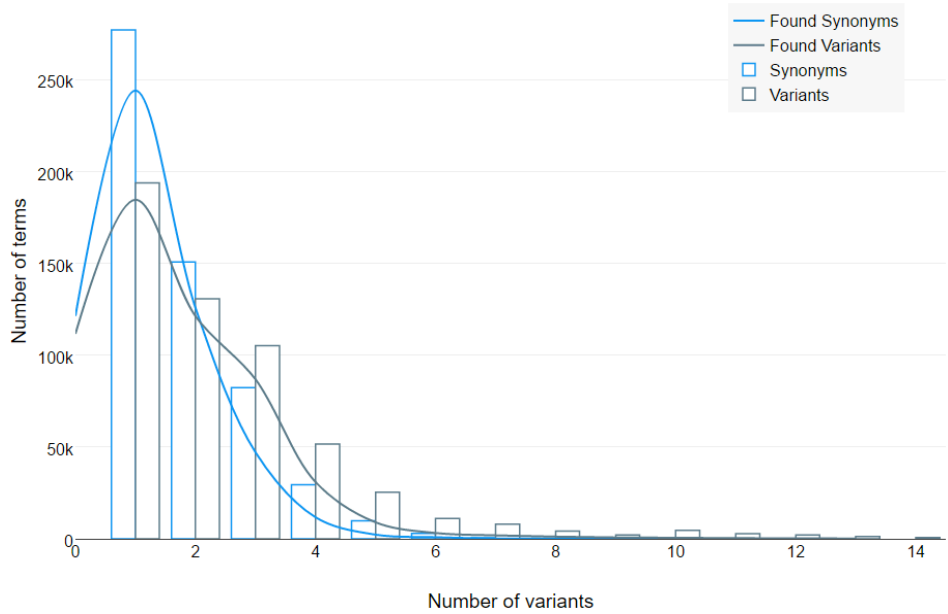

a) Uniprot

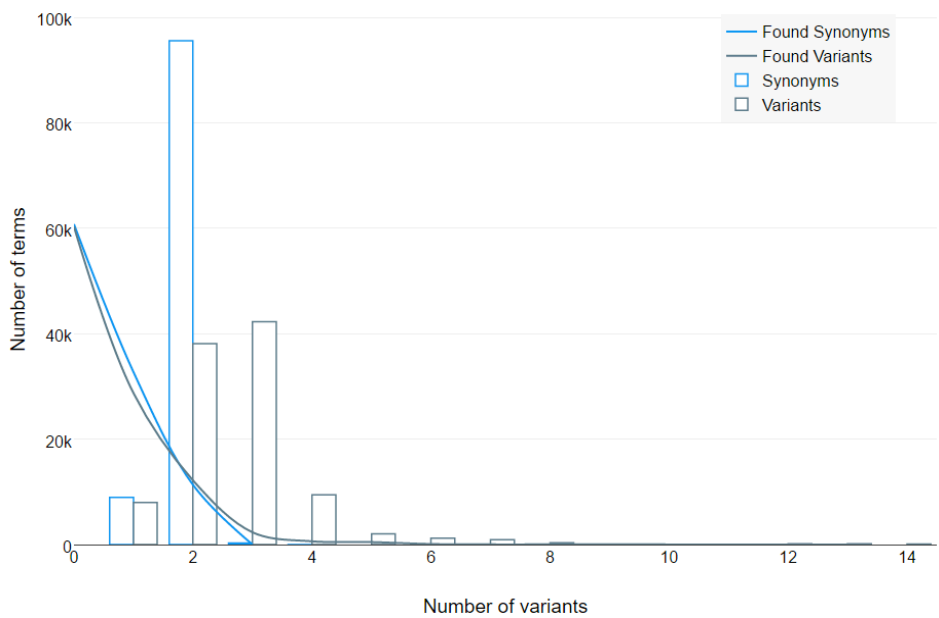

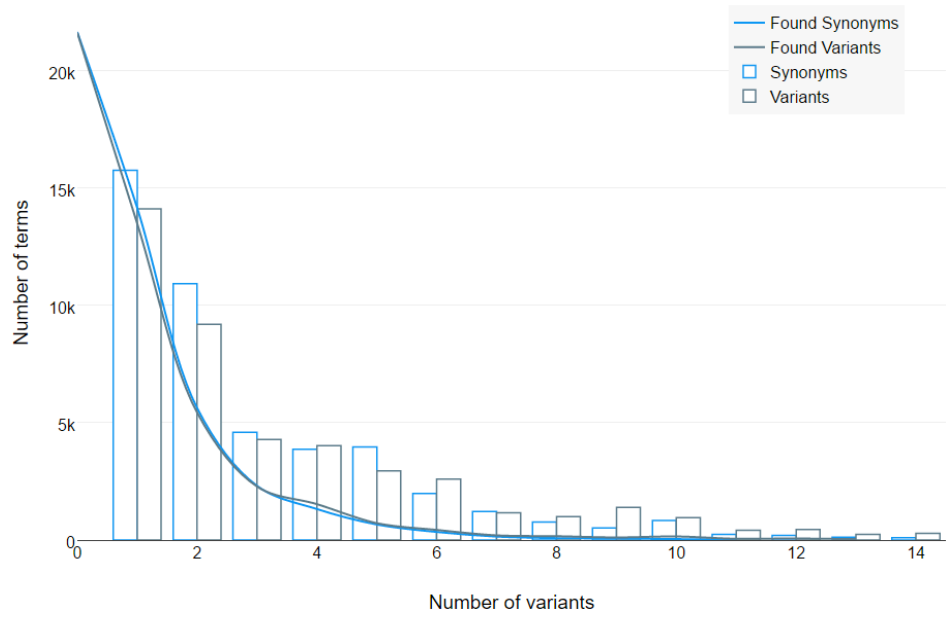

b) ChEBI

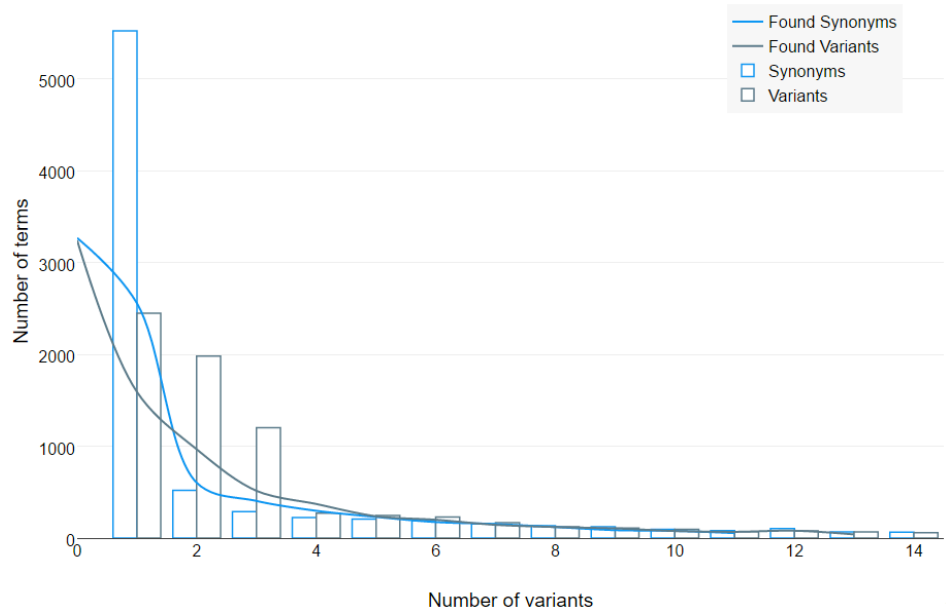

c) GeneOntology

d) DrugBank

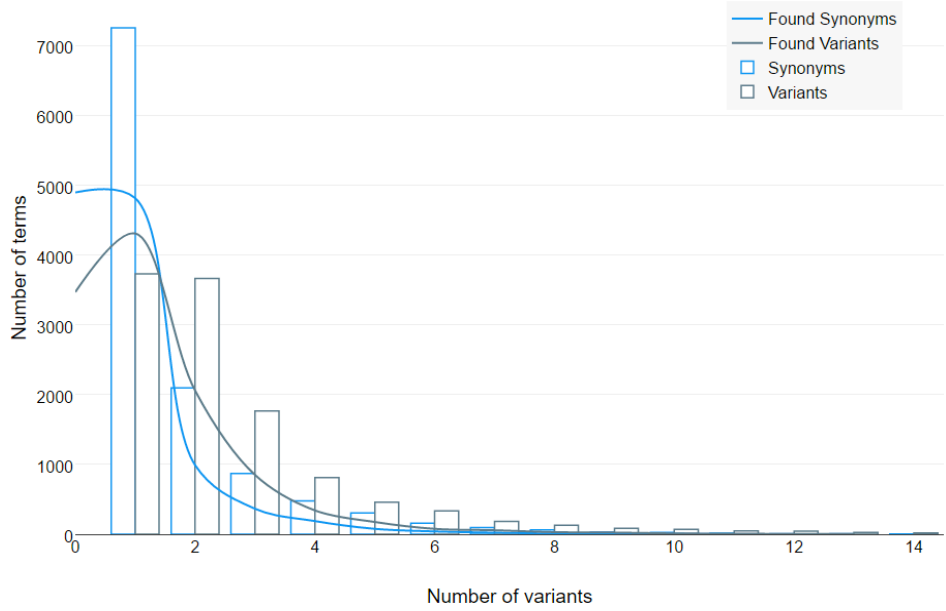

e) ICD10

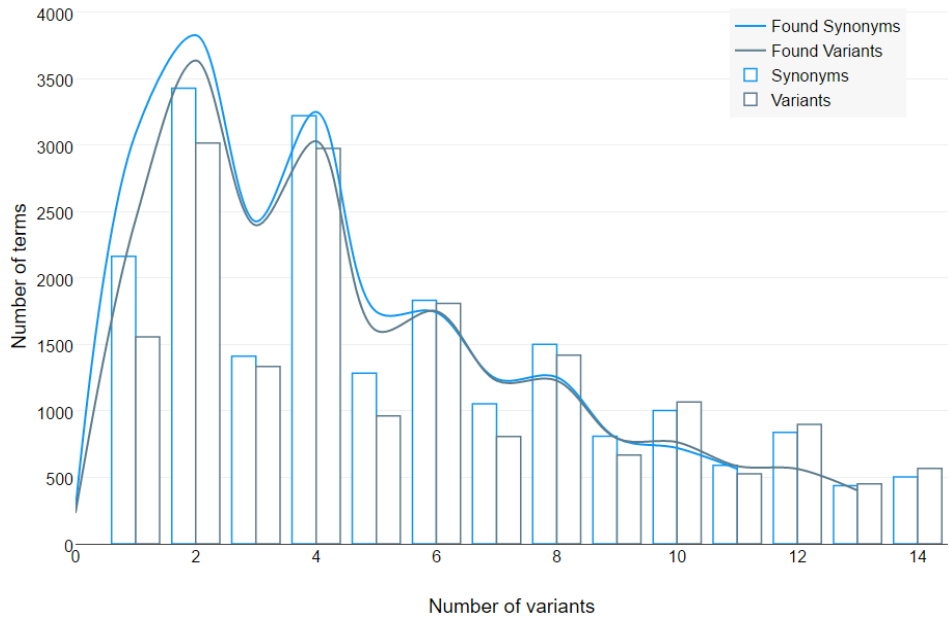

f) MeSH

**Figure 2.** Number of objects that have specific amounts of original synonyms, generated variants and variants found in the papers.

There are original synonyms in the ontologies, generated variants and how much of them were found in the papers. It shows that there are significant chances to found term's spelling that was never mentioned in the ontology. Thus, it reduces recall of the relevant papers.

We were going to use the MSH WSD Data Set [15] initially for the ambiguity testing purposes, but it turned out to contain generic words only. So, we performed a generic wide search across all ontologies and variants to obtain low-level detalization.

There is also a case of "artificial" ambiguity in ontologies. It is caused by intersection of alternative names and terms' descriptions in attempt of extending variants to increase recall. "Carbon monoxide" example is provided in the Table 3.

**Table 3**. Example of the objects with the same spelling.

| Primary name | Ontology | Matched Objects | Synonyms |
|---|---|---|---|
| Carbon monoxide | ICD10 | Accidental poisoning by and exposure to other gases and vapours (X47) | utility gas, utility ga, helium, motor exhaust ga, accidental poisoning by and exposure to other gases and vapours, sulfur dioxide, lacrimogenic gas, **carbon monoxide**, motor exhaust gas, nitrogen oxides |
| | | Intentional self-poisoning by and exposure to other gases and vapours (X67) | utility gas, intentional self-poisoning by and exposure to other gases and vapours, helium, sulfur dioxide, lacrimogenic gas, **carbon monoxide**, motor exhaust gas, nitrogen oxides |
| | | Poisoning by and exposure to other gases and vapours, undetermined intent (Y17) | poisoning by and exposure to other gases and vapours, undetermined intent, utility gas, helium, sulfur dioxide, lacrimogenic gas, **carbon monoxide**, motor exhaust gas, nitrogen oxides |
| | MeSH | Carbon Monoxide (D002248) | **carbon monoxide**, monoxide, carbon |
| | ChEBI | carbon monoxide (CHEBI:17245) | **carbon monoxide**, CO |
| | Drugbank | Carbon monoxide (DB11588) | **carbon monoxide**, lung diffusion test mix nohco, lung diffusion test mixture, lung diffusion test mixture gas, carbon monoxide, compressed air medical G.M., lung diffusion test mix no Ne Co, carbon(II) oxide, Co-HE-O2-N2 mixture, carboneum oxygenisatum, Co-NE-O2-N2 mixture, carbon monox, helium, oxygen, nitrogen L.D.M., CO |
| CH | GeneOntology | naringenin-chalcone synthase activity (GO:0016210) | **CH**, malonyl-coa:4-coumaroyl-coa malonyltransferase(cyclizing), chalcone synthase activity, chalcone synthetase activity, naringenin-chalcone synthase activity, flavanone synthase activity, DOCS, flavonone synthase activity, |

| | | | DOC, CHS |
|---|---|---|---|
| | Uniprot | Cheilanthifoline synthase (C7195_ESCCA) | **CH**, CHS, cytochrome P450 719A5, cheilanthifoline synthase |
| | | Canavanine hydrolase (CANHY_HELVI) | **CH**, canavanine hydrolase |
| | ChEBI | methanylylidene group (CHEBI:29432) | **CH**, methanylylidene group |
| | | methylidyne group (CHEBI:29429) | **CH**, methylidyne group |
| | | methanetriyl group (CHEBI:29433) | **CH**, methanetriyl group |
| | Drugbank | N-Cyclohexyltaurine (DB03309) | CHES, n-cyclohexyltaurine, **CH** |
| IMP | MeSH = ChEBI | Inosine Monophosphate (D007291) | ribosylhypoxanthine monophosphate, inosinic acid, **IMP**, inosinate, sodium, sodium inosinate, inosine monophosphate, acids, inosinic, monophosphate, ribosylhypoxanthine, inosinic acids, monophosphate, inosine, acid, inosinic |
| | DrugBank | Imipenem (DB01598) | imipemide, imipenem anhydrou, imipenem anhydrous, n-formimidoylthienamycin, imipenem, **IMP**, imipenem and cilastatin for injection, USP, ran-imipenem-cilastatin, imipenem, n-formimidoyl thienamycin, imipenem and cilastatin, imipenemum, imipenem and cilastatin for injection, -USP, primaxin 250, imipenem and cilastatin for injection USP, (5R,6S)-6-((R)-1-Hydroxyethyl)-3-(2-(iminomethylamino) ethylthio)-7-oxo-1-azabicyclo(3.2.0) hept-2-ene-2-carbonsaeure, primaxin IV 500, primaxin 500, primaxin IV 250/250 add-vantage vial, imipenem and cilastatin for injection, usp, imipenem and cilastatin for injection,-usp, (5R,6S)-3-(2-formimidoylamino-ethylsulfanyl)-6-((R)-1-hydroxy-ethyl)-7-oxo-1-aza-bicyclo[3.2.0] hept-2-ene-2-carboxylic acid, imipenem and cilastatin for injection, tienamycin, imipenem and cilastatin for injection usp, imipenem and cilastatin for injection-USP, imipenem and cilastatin for injection-usp, primaxin IV, primaxin-iv, n-formimidoyl thienamycin, (5R,6S)-3-((2-(formimidoylamino) ethyl) thio)-6-((R)-1-hydroxyethyl)-7-oxo-1-azabicyclo(3.2.0) hept-2-ene-2-carboxylic acid |
| | GeneOntology. The same | obsolete mitochondrial inner | **IMP**, obsolete mitochondrial inner membrane peptidase activity, mitochondrial inner membrane |

| | | membrane peptidase activity (GO:0004244) | peptidase activity |
| | | mitochondrial inner membrane peptidase complex (GO:0042720) | **IMP**, mitochondrial inner membrane peptidase complex |
| | ChEBI | IMP (CHEBI:17202) | **IMP**, C10H13N4O8P |
| | Uniprot. The same for various organisms | Inositol monophosphatase (IMPA1_DICDI) | IMPase, **IMP**, inositol-1(or 4)-monophosphatase, inositol monophosphatase, d-galactose 1-phosphate phosphatase |
| | | Inositol monophosphatase (IMPP_MESCR) | IMPase, **IMP**, inositol-1(or 4)-monophosphatase, inositol monophosphatase |
| | | Inositol monophosphatase ttx-7 (IMPA1_CAEEL) | IMPase, **IMP**, inositol monophosphatase ttx 7, abnormal thermotaxis protein vii, inositol-1(or 4)-monophosphatase, abnormal thermotaxis protein-7, inositol monophosphatase ttx vii, inositol monophosphatase ttx-vii, abnormal thermotaxis protein7, d-galactose 1-phosphate phosphatase, abnormal thermotaxis protein 7, inositol monophosphatase ttx-7, abnormal thermotaxis protein-vii, inositol monophosphatase ttx7 |

This leads to the necessity of human validation of the same concepts identification. Such functionality exists in sci.AI to validate several ID's from the various ontologies for the same term (Fig. 3).

## 4. Decreasing retrieving uncertainty with the precise semantic labeling feature of the sci.AI platform

Formalization and statistics above show that uncertainty is not an exception but basic feature of the biomedical text mining. This uncertainty might lead to significant deviations when interpreting academic papers with unsupervised methods only. While it might be acceptable for fiction literature mining, because the major task there is context and sentiments analysis that acts as a smoothing function—such uncertainty might contradict goals of mining STEM research communication, where we are looking for the anomalistic or novel discoveries, exact objects interactions, verification of facts, and relations between statements in various texts. This is why accepting uncertainty might have significant negative consequences on the LBD.

**Figure 3.** Labeling term with several objects.

In order to address this issue, we implemented the sci.AI platform that has supervised labeling functionality on top of the text mining framework. After initial automatic terms recognition, no matter whether precision is 70% or 99%, users can make final verifications to level up recognition precision to 100%. From the perspective of the search and text mining algorithms, this means removing any uncertainty, which, in turn, leads to exact papers extraction in an SQL-like querying manner. Thus, assuming that author will always label terms correctly, maximum precision and recall will be achieved.

Human-made corrections will be used as training data for the next processings of a text. Such learning with human feedback provides steady path to gradient growth of text mining quality.

Current version of the sci.AI allows to upload text, then performs Named Entities Recognition (NER) task automatically. Author or annotator can validate labeling results via interface and export final structured text to the XML file (Fig.4).

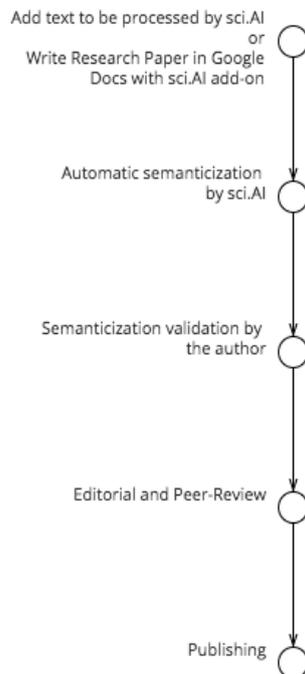

**Figure 4.** Paper semanticization in sci.AI.

Development roadmap includes release of the next features:
- Machine Learning based analysis to provide the most likeable variants in the first place. This feature will be based on the (a) logs of the terms validation events and (b) statistical co-occurrences of the terms in all available texts. We expect that it will provide approximately 90% recognition rate, as reported by NN researchers [7, 8]. Introduction of the NN prioritisation is expected to reduce cases of the necessary authors intervention to the reasonable minimum. Internal time tracking done by our team members suggest that 10 pages validation time will be reduced from 1 h to 15 min, approximately. Current statistics will be corrected after actual feature release and making more measures for the various annotators and texts with different density of the terms;
- Graph based WSD connections between objects and existing metaontologies data;
- Generating JATS, RDF/XML and RDFa files;
- Validation by the readers, not only by the authors;
- Concepts interactions labeling, for example, protein-protein.

The system can be embedded into the publishing process directly. Both authors and editors can create semanticized versions during submission of even publish new version of the digital paper. Key features of the current production version are as follows:

1. Automated metatagging of biomedical concepts, (named-entity recognition, with further context-dependant semanticization of terms). Current tags contain links to the related objects in the ontology;
2. User-friendly web preprint for tags editing and recognition supervising;
3. Web application for easy integration into the existing publishing process.

sci.AI allows labeling a term with several term-2-ontologyId relationships. For example, '*serotonin*' is the object UID=*D012701* in *MeSH* and UID=*28790* in *ChEBI* ontologies simultaneously. Author or reader can suggest additional UIDs too or can correct existing one. This functionality contributes to the global connectivity between terms and might support UMLS Word Sense Disambiguation (WSD) works.

We expect another possible positive effect for the future papers too. Authors might come to the same single spelling variant of the concept, like "ClNa" only and not "salt" or "NaCl".

As of the end of 2016, sci.AI is in the first phases of its long-term development roadmap from being a semanticization tool to becoming a full-fledged artificial intelligence (AI) tool applied to the life sciences. Wide adoption of this application will extend the publisher's role even further into research results delivery to the intended target audience.

**5. Discussion**

This paper is the first in our series of researchers about precise semantic labelling of life sciences texts. Our goal was to focus on dependency of the paper's influence on two fundamental factors: ambiguity and variability of terms. In order to avoid excessive complication, we made several assumptions which may bias the results. These simplifications will be addressed in follow-up studies:

1. Prior precision estimation is calculated with assumption that probability of retrieving a paper with concept $c_i$ when searching for ambiguous term $t_i$ might have uniform distribution. In fact, it has a nonlinear distribution, as shown in statistics in Tables 1, 2 and Fig. 2.

2. Prior recall estimation is calculated with assumption that probability of retrieving a paper with concept $c_i$ when searching for term $t_i$ with multiple variants might have uniform distribution. In fact, it has a nonlinear distribution as shown in statistics in Table 1, 2 and Fig. 2.

3. Simplified dependency of recall from ambiguity and precision from variability.

4. Categories of terms variability and implementation of the terms variant generator deserve full comprehensive description in the following research.

6. We intended to show fundamental specifics of biomedical language that makes it is challenging to achieve 100% recognition of terms with unsupervised methods only. Still, there are various NLP approaches including metaontologies like UMLS based disambiguation and statistical methods that significantly improve terms recognition. Those methods are integrated by sci.AI development team and performance of each of them will be evaluated in separate paper.

7. We assumed that there is the same number of concepts and objects within single ontology.

8. "Human factor" was removed from consideration by assuming that author can always correctly label every biomedical concept in own manuscript. Under "precise labeling" we mean "labeling verified by the actual text's author".

9. There are several studies, where researchers propose models of the future paper's success, for example, [19]. Future analysis might take into consideration ambiguity and variability as variables in the prediction models.

10. Part of speech tagging might improve precision of the variants validation. This functionality exists in the sci.AI but was not applied for the statistics calculation.

11. We assume that all possible spelling variants were generated. Further validation is required.

**Appendix A. Examples of the objects' synonyms and generated variants**

Variants, that were found in actual papers are marked with * and DOI of one of the retrieved papers.

**Primary term:** Peroxisome proliferator-activated receptor gamma coactivator 1-alpha

**Ontology:** Uniprot [PRGC1_HUMAN]
**Synonyms:** PGC-1-alpha, PPAR-gamma coactivator 1-alpha, PPARGC-1-alpha, Ligand effect modulator 6
**Variants:** *PGC-1alpha [10.1186/1750-1326-4-10], PPARGC 1-alpha, *PPAR γ-coactivator 1α [10.1038/nutd.2011.3], *peroxisome proliferator-activated receptor-γ-coactivator 1-α [10.1016/j.molmet.2015.09.003], PPARGC-i-α, *PPAR-gamma coactivator 1α [10.1186/1476-511X-10-246], *PPARGC1-α [10.1186/1743-7075-7-88], *PPAR-gamma coactivator 1alpha [10.1155/2008/418765], *peroxisome proliferator-activated receptor γ coactivator 1-alpha [10.1038/srep18011], *peroxisome proliferator-activated receptor-γ-coactivator 1α [10.1210/me.2014-1164], *PPAR γ coactivator 1 α [10.1074/jbc.M115.636878], *peroxisome proliferator-activated receptor γ coactivator 1-α [10.1016/j.molmet.2015.09.003], PGC-i-α, *PGC-1 α [10.1038/ncomms10210], *PPAR-γ-coactivator 1-alpha [10.1371/journal.pone.0055940], *PGC-1α [10.7554/eLife.03245], *ligand effect modulator-6, PPARGC i-α, *PPARGC-1 α [10.1074/jbc.M113.512483], PPARGC ialpha, *PPAR-gamma coactivator 1 α [10.3892/mmr.2013.1714], *PPAR-γ coactivator 1 α [10.1074/jbc.M115.636878], PPAR γ coactivator 1alpha, *ligand effect modulator 6, PGC-i alpha, *peroxisome proliferator-activated receptor-γ-coactivator 1 alpha [10.1038/srep18011], *PPAR-γ coactivator 1α [10.1038/nutd.2011.3], *PPARGC 1alpha [10.1038/nm.2049], PPARGC-ialpha, *peroxisome proliferator-activated receptor γ coactivator 1alpha [10.1111/jnc.12089], PGC iα, ligand effect modulator6, *peroxisome proliferator-activated receptor gamma coactivator 1 alpha [10.1152/ajpgi.00270.2015], *PPARGC 1 α [10.1074/jbc.M113.512483], PGC-i α, *PPAR-γ-coactivator 1 α [10.1074/jbc.M115.636878], *PGC1-α [10.7150/ijbs.7972], *PGC 1α [10.7554/eLife.03245], PPARGC iα, *PPARGC1 alpha [10.1016/j.jnutbio.2009.03.012], *PGC1-alpha [10.2527/jas.2009-1896], *PPARGC1-alpha [10.1016/j.jnutbio.2009.03.012], *PGC1α [10.1016/j.molmet.2015.08.002], *peroxisome proliferator-activated receptor gamma coactivator 1-alpha [10.1152/ajpgi.00270.2015], *PPAR-γ coactivator 1 alpha [10.1371/journal.pone.0055940], ligand effect modulator vi, *peroxisome proliferator-activated receptor γ coactivator 1 alpha [10.1038/srep18011], *PPARGC 1α [10.1371/journal.pgen.1005062], PPARGC-i α, PPARGC i-alpha, *peroxisome proliferator-activated receptor gamma coactivator 1alpha [10.1007/s00125-006-0268-6], *peroxisome proliferator-activated receptor gamma coactivator 1α [10.1038/ncomms3906], *PPARGC-1-α [10.1074/jbc.M113.512483], *PPAR-γ coactivator 1-α [10.1074/jbc.M115.636878], PPARGC-i-alpha, *peroxisome proliferator-activated receptor γ coactivator 1 α [10.1016/j.molmet.2015.09.003], PPAR-γ coactivator 1alpha, *PPAR γ-coactivator 1-α [10.1074/jbc.M115.636878], *PGC 1-alpha [10.1038/sj.ijo.0803567], *PGC-1-α [10.1038/ncomms10210], *PPAR γ-coactivator 1-alpha [10.1371/journal.pone.0055940], PPARGC-iα, *PGC1 alpha [10.2527/jas.2009-1896], *peroxisome proliferator-activated receptor γ-coactivator 1-α [10.1016/j.molmet.2015.09.003], *PPARGC 1-α [10.1074/jbc.M113.512483], *PGC 1 α [10.1038/ncomms10210], PGC-i-alpha, PPARGC i α, *PPARGC-1α [10.1371/journal.pgen.1005062], *peroxisome proliferator-activated receptor coactivator 1α [10.1210/me.2014-1164], *PPAR-gamma coactivator 1 alpha [10.1152/japplphysiol.00780.2009], *PPAR-γ-coactivator 1α [10.1038/nutd.2011.3], *PGC 1alpha [10.1186/1750-1326-4-10], *PGC-1 alpha [10.1038/sj.ijo.0803567], *PPAR-γ coactivator 1-alpha [10.1371/journal.pone.0055940], PPARGC-1-alpha, *peroxisome proliferator-activated receptor-γ-coactivator 1alpha [10.1111/jnc.12089], PGC i α, PPARGC i alpha, *PPARGC1 α [10.1186/1743-7075-7-88], *PPAR γ coactivator 1α [10.1038/nutd.2011.3], PGC-iα, PPARGC 1 alpha, *PPAR-γ-coactivator 1-α [10.1074/jbc.M115.636878], *PGC1 α [10.7150/ijbs.7972], *peroxisome proliferator-activated receptor gamma coactivator 1 α [10.3168/jds.2015-9847], *PPARGC1α [10.7554/eLife.18206], *PPAR γ coactivator 1 alpha [10.1371/journal.pone.0055940], *PGC 1 alpha [10.1038/sj.ijo.0803567], PGC i-α, PGC i-alpha, *peroxisome proliferator-activated receptor γ-coactivator 1 α [10.1016/j.molmet.2015.09.003], PPAR-γ-coactivator 1alpha, *peroxisome proliferator-activated receptor-γ-coactivator 1 α [10.1016/j.molmet.2015.09.003], *PPAR γ coactivator 1-alpha [10.1371/journal.pone.0055940], *PPARGC1alpha [10.1186/s13395-016-0083-9], ligand effect modulator-vi, *PGC-1-alpha [10.1038/sj.ijo.0803567], *peroxisome proliferator-activated receptor γ-coactivator 1α [10.1210/me.2014-1164], PGC i alpha, *PPAR-gamma coactivator 1-α [10.3892/mmr.2013.1714], PPAR γ-coactivator 1alpha, PGC ialpha, PPARGC-i alpha, *PPAR γ coactivator 1-α [10.1074/jbc.M115.636878], *PGC 1-α [10.1038/ncomms10210], *peroxisome proliferator-activated receptor gamma coactivator 1-α [10.3168/jds.2015-9847], *PPAR γ coactivator 1 alpha [10.1371/journal.pone.0055940], PPARGC-1 alpha, *PPAR γ-coactivator 1 α [10.1074/jbc.M115.636878], *PPAR-gamma coactivator 1-alpha [10.1152/japplphysiol.00780.2009], *peroxisome proliferator-activated receptor γ-coactivator 1alpha [10.1111/jnc.12089], *PGC1alpha [10.1677/jme.1.01499], *peroxisome proliferator-activated receptor γ-coactivator 1-alpha [10.1038/srep18011], PGC-ialpha, *PPAR-γ-coactivator 1 alpha [10.1371/journal.pone.0055940], *peroxisome proliferator-activated receptor γ-coactivator 1 alpha [10.1038/srep18011], *PPARGC-1alpha [10.1038/nm.2049], *peroxisome proliferator-activated receptor-γ-coactivator 1-alpha [10.1038/srep18011]

**Primary term:** Interleukin-1 receptor type 2
**Ontology:** Uniprot [IL1R2_HUMAN]
**Synonyms:** IL-1R-2, IL-1RT-2, IL-1RT2,CD121 antigen-like family member B, CDw121b, IL-1 type II receptor, Interleukin-1 receptor beta, Interleukin-1 receptor type II
**Variants:** interleukin1 receptor type 2, *interleukin-i receptor type-ii [10.1021/ac800928z], *interleukin 1 receptor type-ii [10.1021/ja043466g], *interleukin i-receptor type II [10.1021/ac800928z], interleukin1-receptor type ii, *interleukin-1 receptor type-ii [10.1021/ja043466g], *interleukin-i-receptor type II [10.1021/ac800928z], interleukin-i receptor type-2, interleukin1 receptor type2, interleukin1 receptor type-ii, interleukin-1-receptor type2, *interleukin-1 receptor type ii [10.1021/ja043466g], *interleukin i-receptor type ii [10.1021/ac800928z], interleukin1 receptor type-2, *interleukin i receptor type-ii [10.1021/ac800928z], interleukin1-receptor type II, interleukin-i receptor type 2, *interleukin-i-receptor type-ii [10.1021/ac800928z], interleukin i receptor type 2, *interleukin 1-receptor type ii [10.1021/ja043466g], *interleukin-1-receptor type-ii [10.1021/ja043466g], interleukin1-receptor type-ii, *interleukin 1-receptor type II [10.1021/ja043466g], interleukin1 receptor type ii, interleukin1-receptor type 2, interleukin-i-receptor type2, *interleukin-1 receptor type 2 [10.1038/mi.2015.108], interleukin1 receptor type II, *interleukin i receptor type II [10.1021/ac800928z], interleukin-i-receptor type-2, *interleukin i receptor type ii [10.1021/ac800928z], interleukin1receptor type-ii, interleukin 1-receptor type2, interleukin-i-receptor type 2, *interleukin-1-receptor type II [10.1021/ja043466g], *interleukin-1 receptor type2, interleukin1receptor type-2, interleukin 1 receptor type2, interleukin1receptor type 2, *interleukin-1-receptor type ii [10.1021/ja043466g], *interleukin 1 receptor type ii [10.1021/ja043466g], *interleukin 1 receptor type-2 [10.1038/mi.2015.108], interleukin i receptor type-2, *interleukin-i-receptor type ii [10.1021/ac800928z], *interleukin-1-receptor type-2 [10.1038/mi.2015.108], *interleukin-1 receptor type-2 [10.1038/mi.2015.108], interleukin1-receptor type2, interleukin i-receptor type 2, *interleukin 1-receptor type 2 [10.1038/mi.2015.108], interleukin1-receptor type-2, interleukin1receptor type ii, interleukin i receptor type2, interleukin i-receptor type-2, interleukin1receptor type2, *interleukin 1 receptor type II [10.1021/ja043466g], *interleukin 1 receptor type 2 [10.1038/mi.2015.108], *interleukin-1-receptor type 2 [10.1038/mi.2015.108], interleukin1receptor type II, *interleukin1receptor type-2 [10.1038/mi.2015.108], *interleukin-i receptor type ii [10.1021/ac800928z], *interleukin-i receptor type II [10.1021/ac800928z], *interleukin i-receptor type-ii [10.1021/ac800928z], *interleukin-1 receptor type II [10.1021/ja043466g], interleukin i-receptor type2, *interleukin 1-receptor type-ii [10.1021/ja043466g], interleukin-i receptor type2, interleukin 1-receptorβ, CD121 antigen-like family member B, interleukin i-receptor-beta, interleukin1 receptor type-II, *interleukin-i receptor type-II [10.1021/ac800928z], interleukin-i-receptor beta, interleukin i receptor-beta, interleukin 1-receptor-β, interleukin i-receptorβ, interleukin1receptor-β, interleukin1receptor type-II, *interleukin-1-receptor-beta [10.1186/1471-2164-11-545], interleukin i receptor β, IL1type II receptor, interleukin-1 receptorβ, interleukin-1 receptor-β, *IL-1RT2 [10.3389/fnint.2013.00061], interleukin 1 receptor-β, *interleukin 1-receptor-beta [10.1186/1471-2164-11-545], interleukin1-receptor beta, interleukin 1-receptorbeta, *IL 1-type II receptor [10.1084/jem.20020906], *IL-1R2 [10.1038/mi.2015.108], interleukin1receptor-beta, *IL-1R-2, interleukin-1-receptor β, interleukin-i-receptor-beta, *IL-1RT ii, interleukin i receptor beta, *IL-1-type II receptor [10.1084/jem.20020906], *interleukin i receptor type-II [10.1021/ac800928z], interleukin 1-receptor β, *interleukin-1 receptor beta [10.1186/1471-2164-11-545], *interleukin 1-receptor beta [10.1186/1471-2164-11-545], interleukin-1-receptorβ, IL i-type II receptor, interleukin 1 receptorβ, *interleukin-i-receptor type-II [10.1021/ac800928z], interleukin i-receptor beta, interleukin1-receptor β, *IL1-type II receptor [10.3389/fcell.2016.00072], interleukin-i-receptorβ, interleukin-i receptor-β, *IL1 type II receptor [10.3389/fcell.2016.00072], *IL-1RT-ii, interleukin i-receptor β, interleukin1-receptor-beta, interleukin-i-receptor-β, interleukin-1 receptor β, interleukin-1 receptorβ, interleukin1-receptor type-II, interleukin-1-receptor-β, *CDw121b, interleukin 1 receptorbeta, interleukin1 receptorbeta, interleukin-i receptor beta, interleukin1 receptorβ, interleukin-1-receptorbeta, interleukin1-receptor-β, interleukin i receptorbeta, CD121 antigen-like family member-b, *IL-1R-ii [10.1177/039463200601900204], *IL-1R 2, interleukin1 receptor-β, interleukin i receptor-β, IL-1RT-2, *interleukin i-receptor type-II [10.1021/ac800928z], IL-i type II receptor, interleukin1-receptorβ, interleukin-i receptorbeta, interleukin-i-receptorbeta, *interleukin 1 receptor beta [10.1186/1471-2164-11-545], interleukin i-receptorbeta, interleukin 1 receptor β, interleukin1 receptor beta, *IL 1 type II receptor [10.1084/jem.20020906], *interleukin 1 receptor type-II [10.1021/ja043466g], *interleukin-1 receptor type-II [10.1021/ja043466g], IL-i-type II receptor, *IL-1R ii [10.1177/039463200601900204], interleukin i receptorβ, interleukin1receptorβ, *interleukin-1-receptor type-II [10.1021/ja043466g], *IL-1 type II receptor [10.1084/jem.20020906], interleukin i-receptor-β, IL-1RT 2, interleukin-i-receptor β, *interleukin 1 receptor-beta [10.1186/1471-2164-11-545], interleukin1 receptor-beta, interleukin-i receptor β, interleukin1receptor β, IL i type II receptor, interleukin1receptorbeta, *interleukin 1-receptor type-II [10.1021/ja043466g], *interleukin-1 receptor-beta [10.1186/1471-2164-11-545], interleukin1receptor beta, interleukin-i receptor-beta, interleukin-i receptorβ, interleukin1-receptorbeta, *interleukin-1-receptor beta [10.1186/1471-2164-11-545]